\documentclass[12pt]{elsarticle}
\usepackage{amsmath}
\usepackage{xcolor}
\usepackage{stackengine}
\usepackage{braket}
\usepackage{bbold}
\usepackage{subcaption}
\usepackage{txfonts}

\usepackage[a4paper, total={7in, 10in}]{geometry}
\title{Increasing the secret key rates and point-to-multipoint extension for experimental coherent-one-way quantum key distribution protocol}
 \author{Venkat Abhignan, Mohit Mittal, Aditi Das, Megha Shrivastava}
 \address{Qdit Labs Pvt. Ltd., Bengaluru - 560092, India}
\begin{document}
\makeatletter
\def\ps@pprintTitle{%
  \let\@oddhead\@empty
  \let\@evenhead\@empty
  \let\@oddfoot\@empty
  \let\@evenfoot\@oddfoot
}
\makeatother
\begin{frontmatter}
\begin{abstract}
   Using quantum key distribution (QKD) protocols, a secret key is created between two distant users (transmitter and receiver) at a particular key rate. Quantum technology can facilitate secure communication for cryptographic applications, combining QKD with one-time pad (OTP) encryption. In order to ensure the continuous operation of QKD in real-world networks, efforts have been concentrated on optimising the use of experimental components and effective QKD protocols to improve secret key rates and increase the transmission between multiple users. Generally, in experimental implementations, the secret key rates are limited by single-photon detectors, which are used at the receivers of QKD  and create a bottleneck due to their limited detection rates (detectors with low detection efficiency and high detector dead-time). We experimentally show that secret key rates can be increased by combining the time-bin information of two such detectors on the data line of the receiver for the coherent-one-way (COW) QKD protocol with a minimal increase in quantum bit error rate (QBER, the proportion of erroneous bits). Further, we implement a point-to-multipoint COW QKD protocol, introducing an additional receiver module. The three users (one transmitter and two receivers) share the secret key in post-processing, relying on OTP encryption. Typically, the dual-receiver extension can improve the combined secret key rates of the system; however, one has to optimise the experimental parameters to achieve this within security margins. These methods are general and can be applied to any implementation of the COW protocol. 
\end{abstract}
\end{frontmatter}
\section{Introduction}
Two distant parties, traditionally called Alice and Bob, share a secret key using QKD with composable and unconditional security derived from the laws of quantum physics \cite{RevModPhys.74.145,RevModPhys.81.1301,Lo2014,RevModPhys.92.025002,Pirandola:20}. The security definition for the QKD protocol is generally determined regardless of its practical implementation in order to achieve what is referred to as composable security \cite{RevModPhys.94.025008}. Because of a comparatively easier configuration and implementation, COW QKD \cite{Damien,Stucki:09} has made substantial experimental progress beyond the fundamentally intriguing Bennett-Brassard 1984 QKD \cite{BENNETT20147}. In order to enhance the secure distance beyond 100 km \cite{Stucki_2009,Korzh2015,Malpani2024} and improve the practicality of the protocol \cite{Walenta_2014,Sibson2017,Sibson:17,roberts2017,Dai:20}, COW QKD has undergone potential experimental alterations. However, it was shown recently that all long-distance implementations of this protocol conducted so far are vulnerable against zero-error attacks \cite{PhysRevLett.125.260510,Trényi_2021,Rey-Domínguez_2024}, which is concerning. Further, to rectify this, Ref. \cite{PhysRevApplied.18.064053} proposed to append a “vacuum-tail” pulse after every encoded signal and use a balanced beam splitter for passive basis choice at Bob. This small modification yields a key rate comparable to the known upper bound of standard BB84, demonstrating that COW-QKD can be securely deployed even in very high-loss (long-distance) optical links. Also, additional vacuum decoy states were used as a countermeasure against zero-error attacks, and the improved asymptotic key rates were proposed from the security proof for COW-QKD \cite{Gao:22}. Using the same variation, finite-size effects in these key rates were recently studied because of the limited resources utilized in a practical COW QKD protocol by quantifying the statistical fluctuations \cite{PhysRevResearch.6.013022}. All these security and experimental studies demonstrate that, in practice, the security of the COW protocol guarantees a secure distance of 100 km between the two users \cite{10955416,sciadv.aec2776}. Considering this, we experimentally study how secret key rates can be increased for distances around 100 km without altering the protocol itself.

Furthermore, we are interested in a simple experimental implementation to show how the protocol can be extended to three users by concurrently sharing the key between Alice and the dual Bob modules (Bob 1 and Bob 2, similar to a point-to-multipoint QKD \cite{Townsend1997,Fröhlich2013}). From an information-theoretic standpoint, fundamental upper bounds on secret-key capacities have been established \cite{PhysRevLett.102.050503}. In particular, the exact secret-key capacity of a pure-loss optical channel was determined using the relative entropy of entanglement (PLOB bound) \cite{Pirandola2017}. An upper bound was given for point-to-point channels in terms of the channel’s squashed entanglement (TGW bound) \cite{Takeoka2014}. These results have been extended to multi-user (broadcast) channels \cite{PhysRevA.96.032318,PhysRevLett.119.150501,7438836}. In addition to our point-to-multipoint experimental results, we compared our measured rates with these information-theoretic bounds and conservative upper bounds obtained from a class of possible collective attacks in the presence of an eavesdropper \cite{Branciard_2008}. We particularly considered the collective beam-splitting attack \cite{RevModPhys.81.1301,Bacco2016,PhysRevA.101.032334}, which yields an undetectable bound in the regime where no QBER is introduced with visibility remaining ideal (unperturbed coherence between signals) and therefore provides a stringent benchmark for our dual Bob COW implementation.

The COW protocol and experimental details are described in Sec. 2, along with the results regarding dual detectors implementation for measuring increased key rates in Sec. 2.1. Furthermore, we experimentally show how the protocol was extended to three users by concurrently sharing the key between Alice and the dual Bob modules. The secret keys created between subsystems (Alice and Bob 1), (Alice and Bob 2) are combined and shared using OTP encryption by Alice to form a final secret key between Alice, Bob 1, and Bob 2. We also derive the secure key rate bounds for our experimental parameters obtained from this implementation, considering a collective beam-splitting attack \cite{Branciard_2008}. These discussions are all detailed in Sec. 2.2. 

\section{COW practical implementation}

\begin{figure}
    \centering
\includegraphics[width=0.75\linewidth]{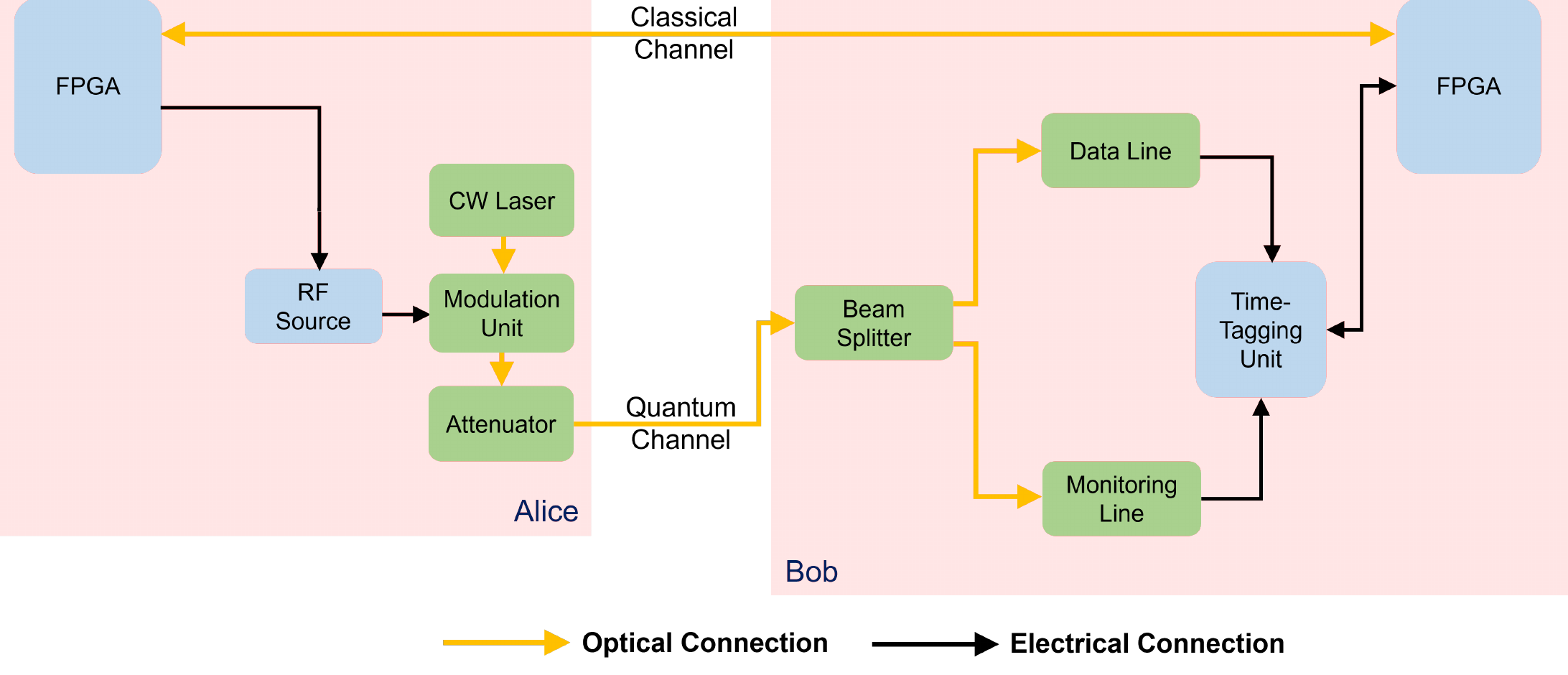}
    \caption{Experimental architecture of COW QKD.}
    \label{fig:enter-label}
\end{figure}

We describe in this section the architecture in a typical practical implementation of COW protocol \cite{Damien} as can be seen in Fig. 1. It consists of a transmitter module (Alice) that uses a continuous wave (CW) laser source and a modulator unit to create a sequence of coherent states, $\ket{0}_t\ket{\sqrt{\mu}}_{t-\tau}$ (two-mode state for bit value 1), $\ket{\sqrt{\mu}}_t \ket{0}_{t-\tau}$ (two-mode state for bit value 0), and $\ket{\sqrt{\mu}}_t\ket{\sqrt{\mu}}_{t-\tau}$ (two-mode state for decoy pulses), where $\mu$ is the mean photon number of the optical pulses. The time between the consecutive pulses is $\tau=1/F$, with $F$ being the repetition rate of the pulses, and $\ket{0}_t$ denotes the vacuum state or no pulse at time $t$. {\it a priori} probabilities $P_{0}=P_{1}=(1-f)/2$ and $P_{\rm decoy}=f$ are used to produce the states for logical bit 0, 1 and the decoy signal, respectively for a given $f$. Here, we generate a sequence of pulses on Alice's side at random using a true random number generator (TRNG).

\begin{figure}[htp]
    \centering
\includegraphics[width=0.85\linewidth]{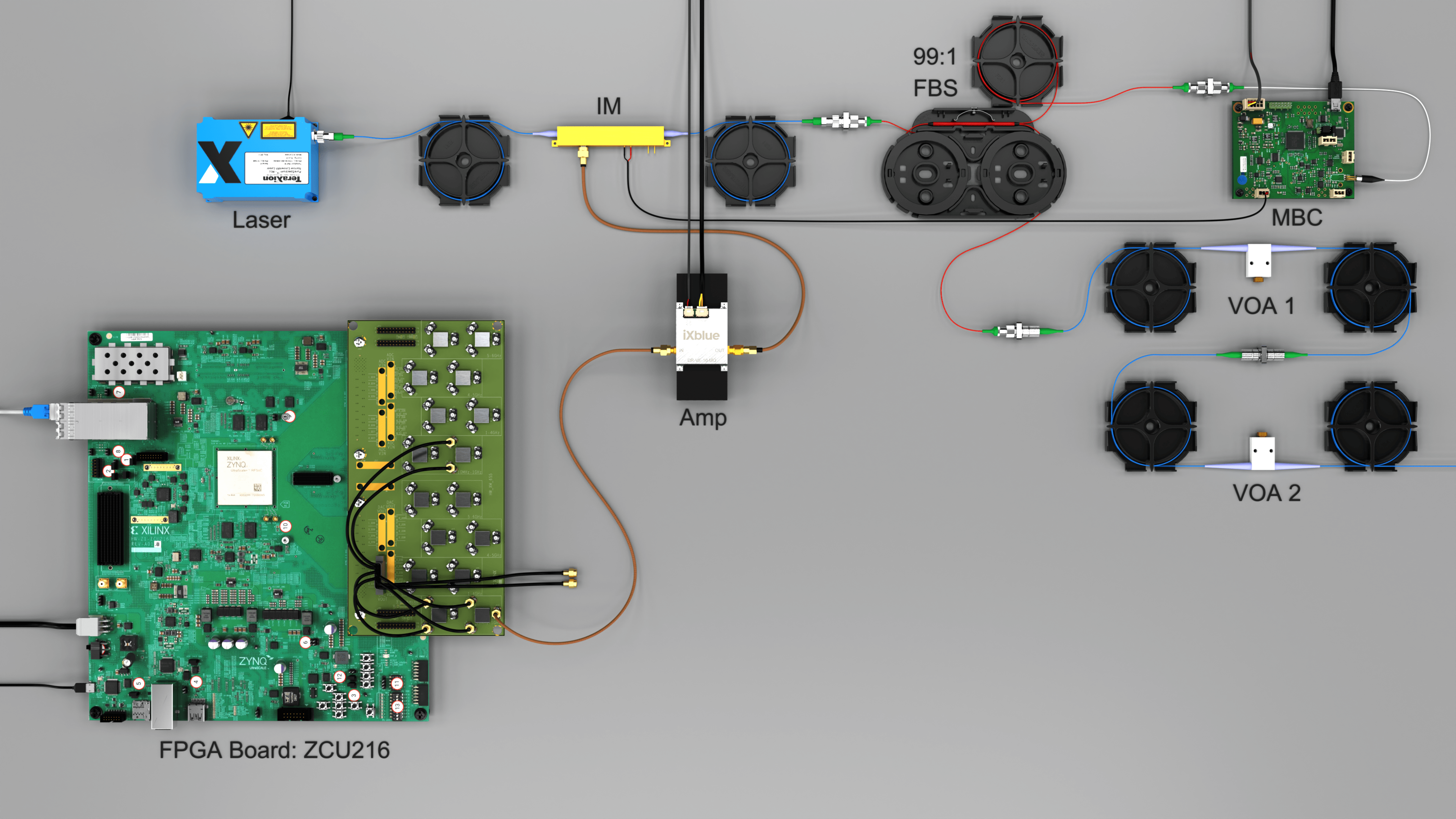}
    \caption{COW QKD: Alice module.}
    \label{fig:enter-label}
\end{figure}

Also, as shown explicitly in Fig. 2, the Alice setup consists of a field programmable gate array (FPGA: ZCU216), which produces RF pulses at the repetition rate of $F = 1$ GHz, which drives an intensity modulator (IM: MXER-LN-10) to produce coherent pulses from the continuous-wave laser signal (PS-NLL-1550.12-080-100-A1) at the same rate. 1\% of the signal is sent to the modulator bias controller (MBC-DG-BOARD-A1) as feedback to IM through a 99:1 fiber beam splitter (FBS) for tuning the bias and stabilizing the operating point of IM. Further, these pulses are attenuated by $\alpha$ (dB) by a set of variable optical attenuators (VOA1 and VOA2) to generate weak coherent pulses with a photon number $\mu$. The average power of these pulses is given by $P_f=\mu F hc/\lambda$ (measured in watts, W) where $h$ is Planck’s constant, $c$ is the speed of light in vacuum, and $\lambda=1550.12$nm is the wavelength of the laser signal. If the initial average power of the pulses generated by the IM is $P_i$ (W), the attenuation $\alpha$ required to reduce the power to $P_f$ is given from 
\begin{equation}
    \alpha=10\log_{10}{\frac{P_f}{P_i}}.
\end{equation}
 
 Alice then transmits these modulated quantum states to the receiver module (Bob) via the quantum channel (fiber), as can be seen explicitly in Fig. 3(a). The fiber causes a fixed amount of loss = $\alpha_d L$ dB (in our experiment, we consider $\alpha_d$=0.22 dB/km). If we are interested in the experiment for a distance of $L=80$ km, a 10 km fiber spool was used (loss of 2.2 dB), and further, to account for 17.6 dB channel loss, an additional 15.4 dB loss was added using attenuators (VOA 1 and VOA 2 in Fig. 2).

\begin{figure}[htp]
\centering
\begin{subfigure}{0.59\linewidth}
\includegraphics[width=\linewidth]{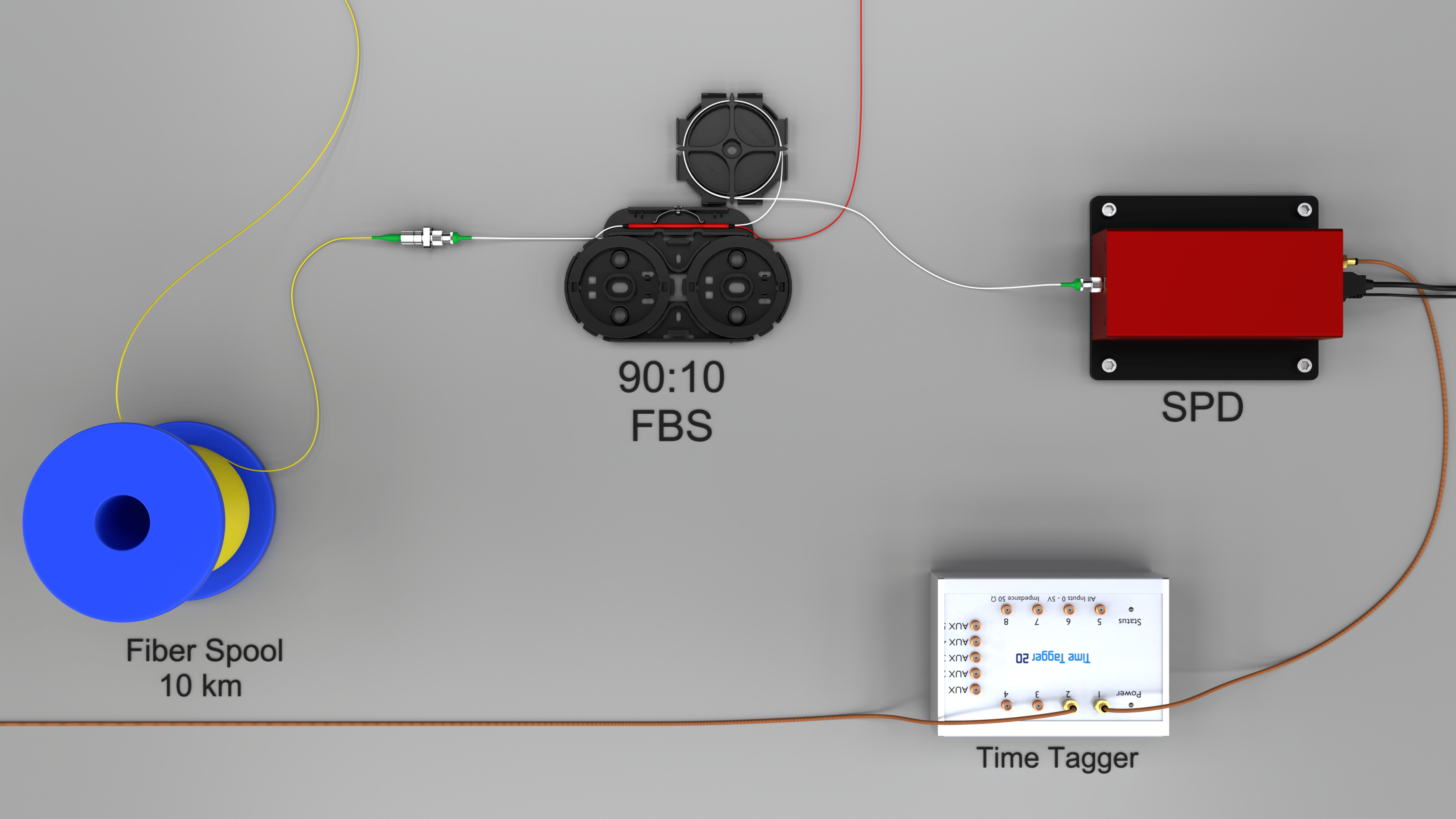}
\caption{Typical Bob module.}
\end{subfigure}
\begin{subfigure}{0.4\linewidth}
\includegraphics[width=\linewidth]{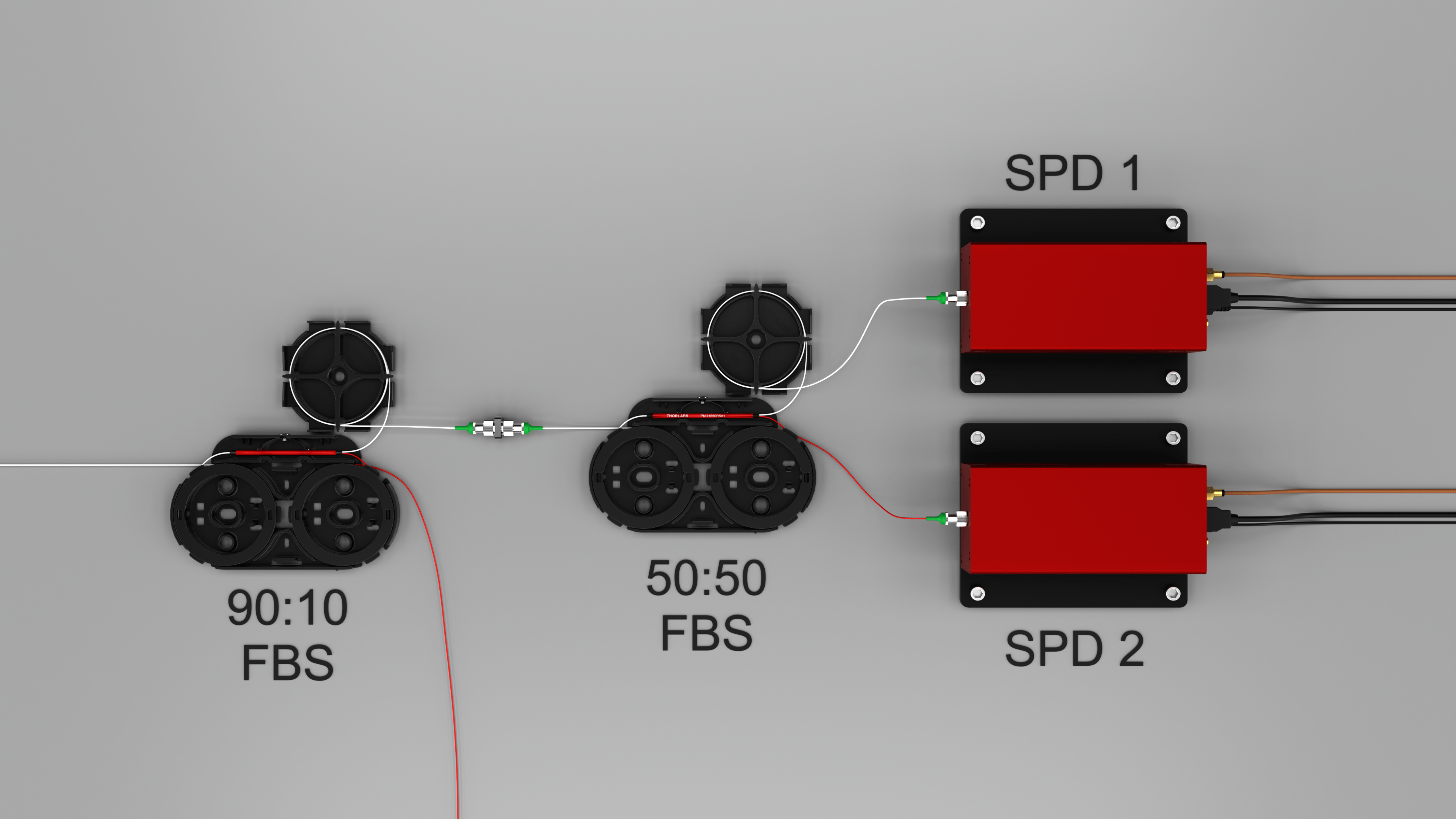}
\caption{Bob module with dual detectors on data line.}
\end{subfigure}
\caption{COW QKD: Bob module.}
\end{figure}

At Bob's end, a 90:10 FBS divides the incoming signals sent into the data (90\%) and monitoring lines (10\%). The monitoring line is usually required to examine any eavesdropping in the quantum channel by analyzing the coherence of nearby non-empty pulses (Visibility). Identifying an eavesdropper-induced breaking of coherence is possible through the monitoring line. However, we are experimenting here without considering the monitoring line since it will remain unperturbed due to our modifications. As shown explicitly in Fig. 3 (a), only the data line is considered after 90:10 FBS. Bob can record information at the single-photon level using single-photon detectors (SPDs). The data line is connected to a SPD (InGaAs/InP Geiger-mode SPD\_OEM\_NIR) with finite quantum efficiency $\eta$ and dead time $t_d$. Dead time of an SPD is the period after a photon is detected, during which the SPD cannot detect any more incoming photons because of quenching of the Geiger-mode avalanche current. This restricts the SPD's detection of the maximum raw counts of photons (rate defined as $C_{exp}$ measured in counts per second, cps). Classically communicating with Alice the detections on the data line and the timing of decoy states (using a time tagger), Bob detects when Alice released a bit state of 0 or 1 and builds a sifted key (rate defined as \textit{SKR} measured in bits per second, bps) from the detected raw photon counts. \textcolor{black}{ As mentioned, the 10\% monitoring arm is unused in Fig.~3. This does not alter the protocol principle: in a full COW deployment, the monitoring arm can be equipped with the usual interferometer and additional SPDs to perform the standard visibility check (As defined in Ref. \cite{Damien}, the visibility is obtained from the detection events at additional SPDs, which serves as the standard metric for monitoring coherence in the COW protocol). Our dual-SPD modification on the data line leaves the monitoring path intact, so the visibility-monitoring function can be introduced without changing the rest of the architecture.}

Ideally, we can theoretically estimate the photon count that SPD can detect $C_{th}^{(t_d\rightarrow0)}$ that is independent of $t_d$ and is only restricted by $\eta$, $\mu$, $F/2$ the initial qubit generation rate, and $\alpha_d$ losses incurred using the fiber of length $L$. This can be determined by \begin{equation}
   C_{th}^{(t_d\rightarrow0)} = \ 0.9 \ \eta \ \mu \ (F/2) \ 10^{\left(\frac{-\alpha_d L}{10}\right)}. \end{equation}
The factor of 0.9 is because the SPD is on the data line after the 90:10 FBS. Further, the theoretical prediction for the raw counts of photons that the SPD can detect $C_{th}$ is confined by $t_d$ which can be quantified from \begin{equation}
    C_{th} = \frac{C_{th}^{(t_d\rightarrow0)}}{1+t_d C_{th}^{(t_d\rightarrow0)}}.
\end{equation}

This expression for $C_{th}$ predicts the $C_{exp}$ best when $C_{th}^{(t_d\rightarrow0)} > 1/t_d$, and it reflects the fact that after each detection, the SPD is temporarily blind due to its dead time $t_d$. While for $C_{th}^{(t_d\rightarrow0)} < 1/t_d$, the impact of $t_d$ on the raw count rate $C_{th}$ is minimal, ensuring that most incident photons are detected without being missed due to $t_d$. In cases where $C_{th}^{(t_d\rightarrow0)} < 1/t_d$, the expression for $C_{th}$ tends to underestimate the actual counts. This is because, at low values of photon flux, most incident photons are detected without being missed due to lower $t_d$. We also typically observed that while $C_{th}^{(t_d\rightarrow0)}$ was in the range of $10^6$ cps while $C_{th}$ was in the range of $10^5$ cps for our most optimal experimental settings which indicates only 10\% of the photons incident on the SPD we used are being recorded. 

\subsection{Doubling secret key rates}

Essentially, to collect more information from the incident photons, $C_{exp}$ can be improved, and consequently \textit{SKR} can be increased by combining the time-bin information of dual SPDs on the data line since primarily $t_d$ of a single SPD causes the bottleneck here. However, we note that this bottleneck exists only with detectors having $C_{th}^{(t_d\rightarrow0)} > 1/t_d$ following the discussion above. These bottlenecks can also be removed using higher detection efficiency and lower dead-time detectors, such as superconducting-nanowire single-photon detectors (ID281 SNSPDs IDQ). However, costly and specialized cryogenic equipment is needed to operate SNSPDs at low temperatures. Further, as shown explicitly in Fig. 3 (b), the data line from the 90\% arm of 90:10 FBS is connected to $1\times2$ 50:50 FBS, and the outputs are linked to SPDs. As discussed previously, it has to be noted that the 10\% arm of the 90:10 FBS used for identifying eavesdroppers remains unperturbed due to dual detectors on the data line. We experimented only with the mean photon number $\mu=0.5$ in this section to show the advantage of dual detectors on the data line. The average power of the optical pulse before attenuation was $P_i=$ 2.49 mW, and $\alpha=$ 75.91 dB attenuation was required as per Eq. 1 to obtain $\mu=0.5$. 

For distances $L=80$ km, $L=100$ km and $L=120$ km the theoretical counts $C_{th}$ (cps) (Eq. (3)), the experimental counts $C_{exp}$ (cps), the sifted key rate \textit{SKR} (bps), the QBER (the ratio of erroneous bits compared to the bits received, occurring due to noise in the quantum channel and SPDs) are shown in Figs. 4, 5, and 6, respectively. The experimental rates $C_{exp}$ and \textit{SKR} are compared using single SPD (1SPD) and dual SPDs (2SPD) in the data line. The SPD efficiency and dead-time can be varied as $\eta=0.15,0.20$ and the range of $t_d=15\muup s-100\muup s$. The $C_{exp}$ of single SPD matches with $C_{th}$ while the behavior of increased $C_{exp}$ of dual SPDs nearly matches with ideal $2 C_{th}$ and the \textit{SKR} increases using dual SPDs. However, as can be seen in the sub-figures (b) and (d) of Fig.s 4, 5, 6, the QBER increases when using dual detectors and is nearly 5-6\% for lower $t_d$, which is the threshold in QKD implementation to detect the presence of a potential eavesdropper in the channel. Also, QBER increases for increasing distance $L$ and is slightly more for $\eta=0.20$ with higher \textit{SKR} than for $\eta=0.15$ with lower \textit{SKR}. It can also be seen that increasing $t_d$ reduced the key rates; however, \textit{SKR} increases when using dual SPDs against using a single SPD. When compared for varying distances $L=80$ km, $L=100$ km, and $L=120$ km, similar behavior is observed where \textit{SKR} remains increased. However, as $L$ increases, the dual detector $C_{exp}$ becomes lesser than ideal $2 C_{th}$. \textcolor{black}{The introduction of a 50:50 FBS prior to detection on the data line has a non-trivial impact on the QBER. While the FBS distributes incident photons equally between the dual SPDs, thereby reducing the per-detector $C$, the dark count rate of each detector remains unchanged, as it is an intrinsic property of the device. Consequently, the signal-to-noise ratio (SNR) in each SPD is degraded because the signal-to-dark count ratio decreases. Given that QBER is directly related to the fraction of erroneous detections, of which dark counts form a dominant contribution, the relative contribution of dark counts to the total detection events increases. This results in a slightly elevated QBER in dual SPDs compared to the single SPD configuration.}

\begin{figure}[htp]
\centering
\begin{subfigure}{0.63\linewidth}
\includegraphics[width=\linewidth]{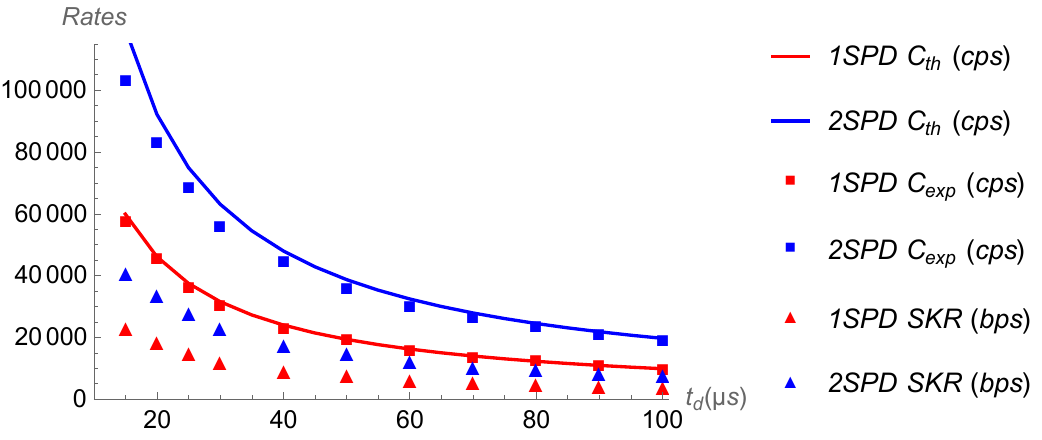}
\caption{Rates when detector efficiency $\eta=0.15$.}
\end{subfigure}
\begin{subfigure}{0.359\linewidth}
\includegraphics[width=\linewidth]{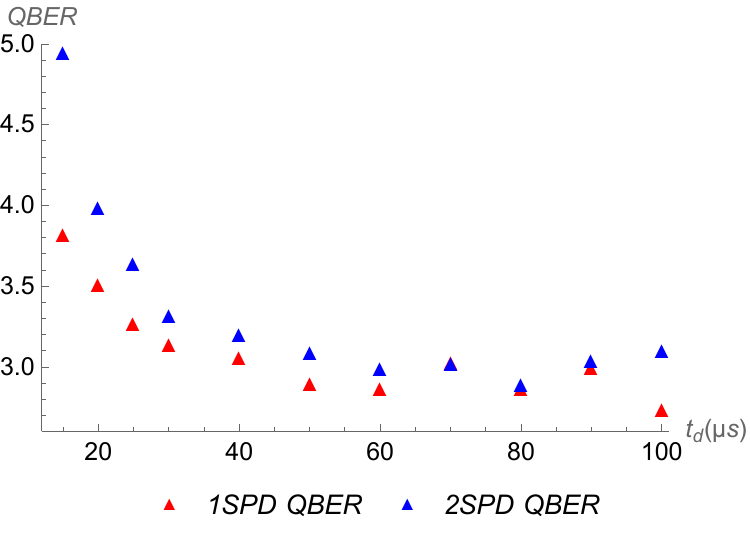}
\caption{QBER when detector efficiency $\eta=0.15$.}
\end{subfigure}
\begin{subfigure}{0.63\linewidth}
\includegraphics[width=\linewidth]{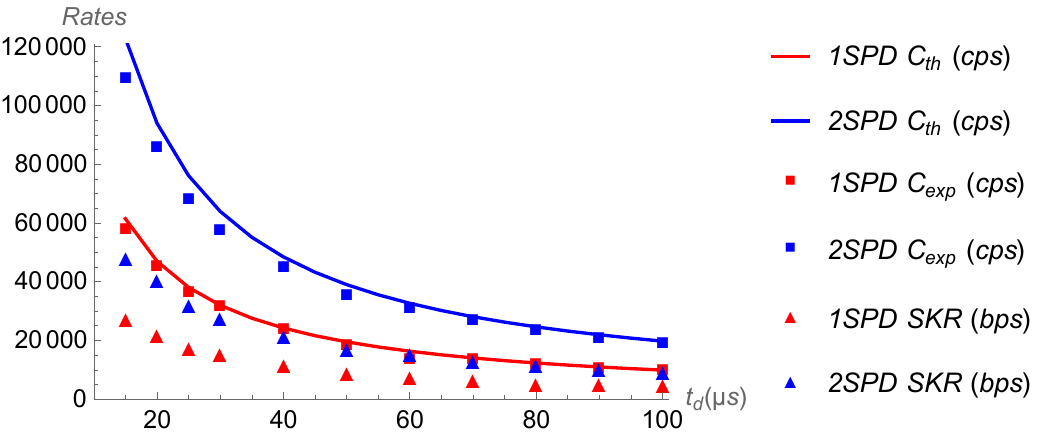}
\caption{Rates when detector efficiency $\eta=0.20$.}
\end{subfigure}
\begin{subfigure}{0.359\linewidth}
\includegraphics[width=\linewidth]{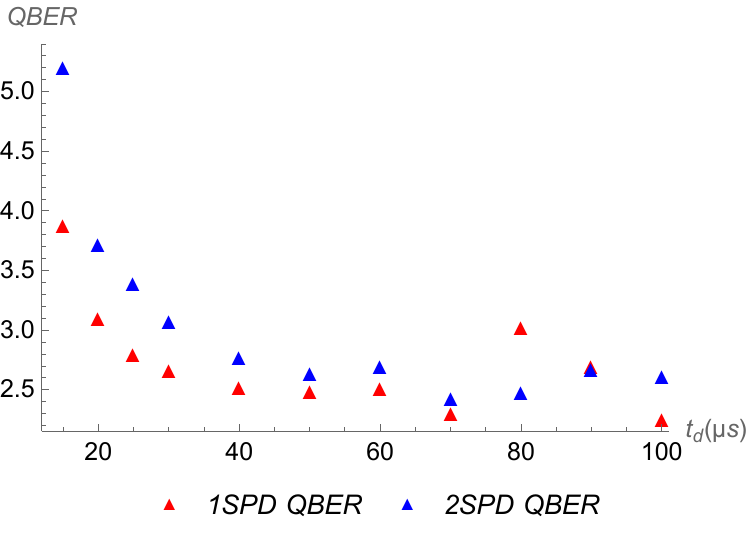}
\caption{QBER when detector efficiency $\eta=0.20$.}
\end{subfigure}
\caption{Comparing the key rates with single detector and dual detectors on the data line for distance $L=80$ km between Alice and Bob.}
\end{figure}

\begin{figure}[htp]
\centering
\begin{subfigure}{0.63\linewidth}
\includegraphics[width=\linewidth]{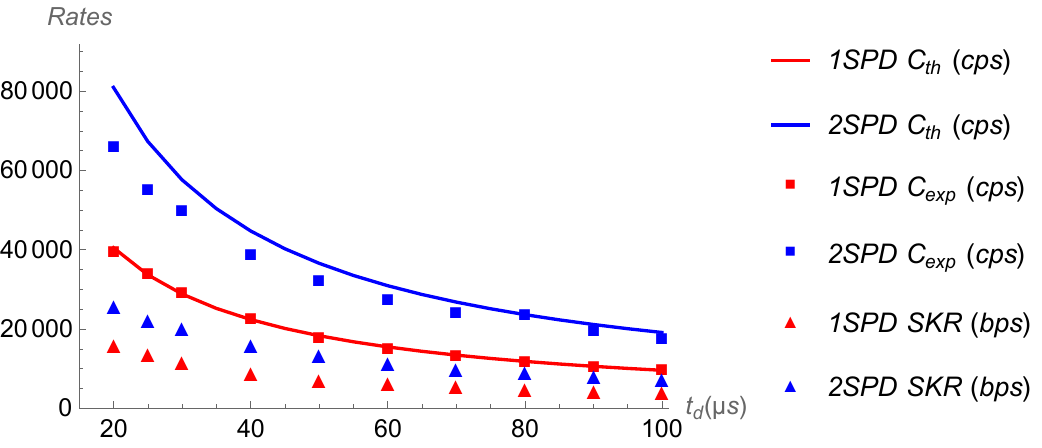}
\caption{Rates when detector efficiency $\eta=0.15$.}
\end{subfigure}
\begin{subfigure}{0.359\linewidth}
\includegraphics[width=\linewidth]{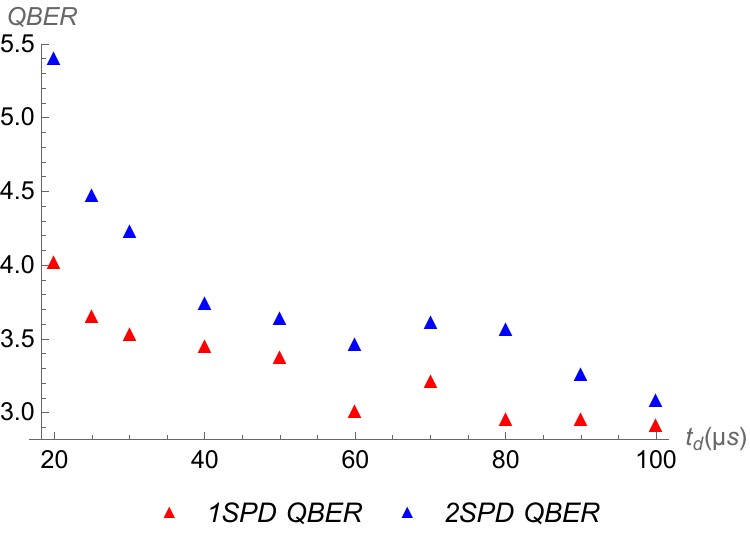}
\caption{QBER when detector efficiency $\eta=0.15$.}
\end{subfigure}
\begin{subfigure}{0.63\linewidth}
\includegraphics[width=\linewidth]{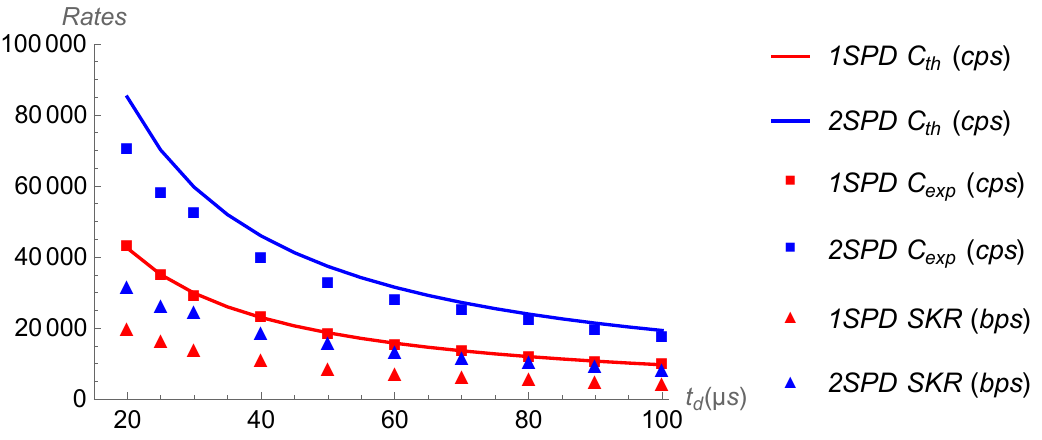}
\caption{Rates when detector efficiency $\eta=0.20$.}
\end{subfigure}
\begin{subfigure}{0.359\linewidth}
\includegraphics[width=\linewidth]{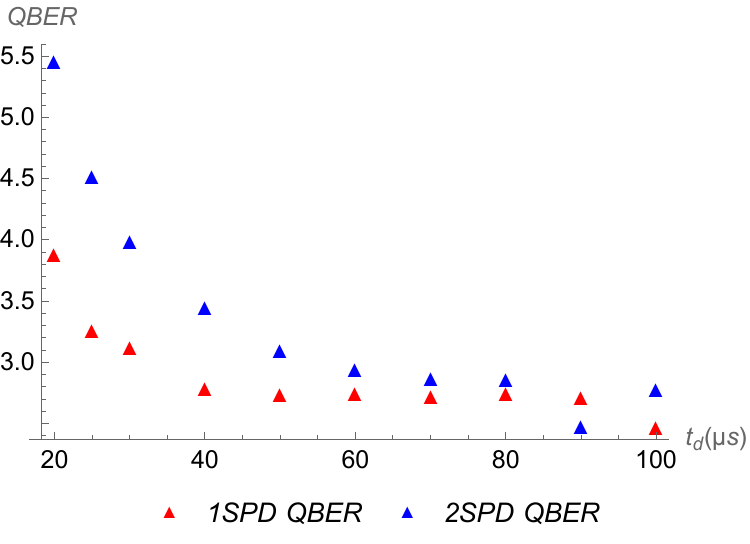}
\caption{QBER when detector efficiency $\eta=0.20$.}
\end{subfigure}
\caption{Comparing the key rates with single detector and dual detectors on the data line for distance $L=100$ km between Alice and Bob.}
\end{figure}

\begin{figure}[htp]
\centering
\begin{subfigure}{0.63\linewidth}
\includegraphics[width=\linewidth]{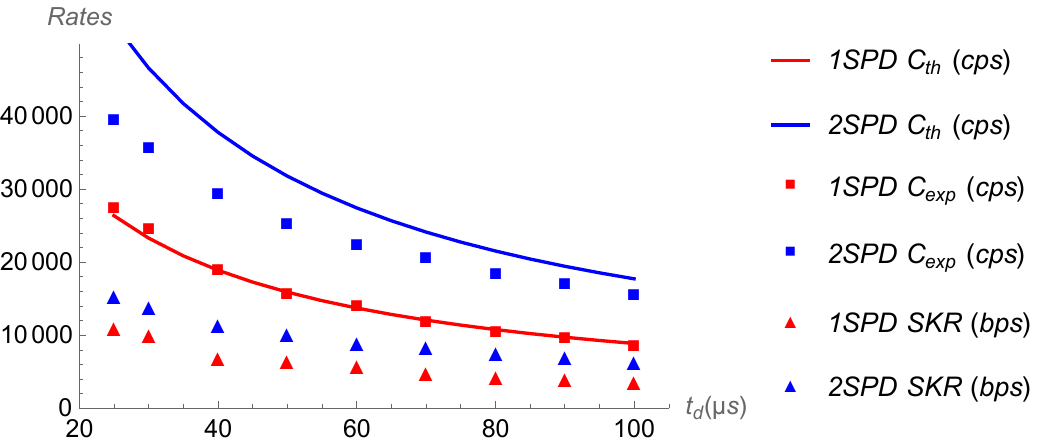}
\caption{Rates when detector efficiency $\eta=0.15$.}
\end{subfigure}
\begin{subfigure}{0.359\linewidth}
\includegraphics[width=\linewidth]{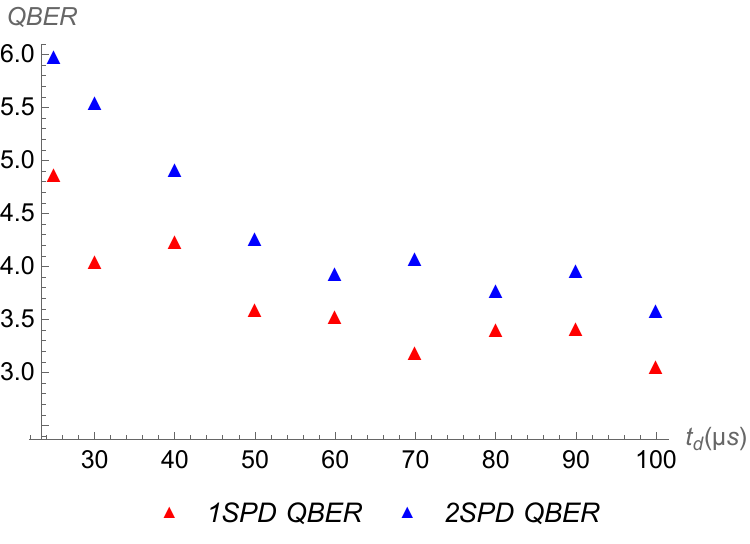}
\caption{QBER when detector efficiency $\eta=0.15$.}
\end{subfigure}
\begin{subfigure}{0.63\linewidth}
\includegraphics[width=\linewidth]{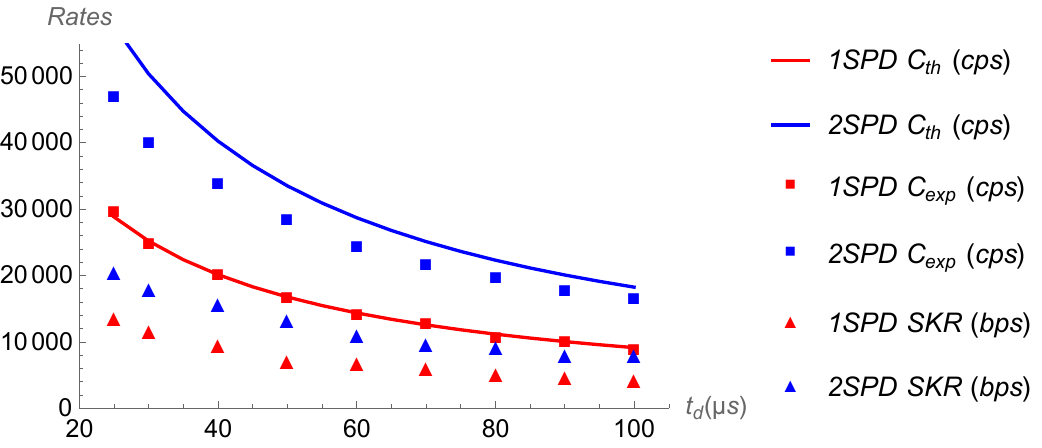}
\caption{Rates when detector efficiency $\eta=0.20$.}
\end{subfigure}
\begin{subfigure}{0.359\linewidth}
\includegraphics[width=\linewidth]{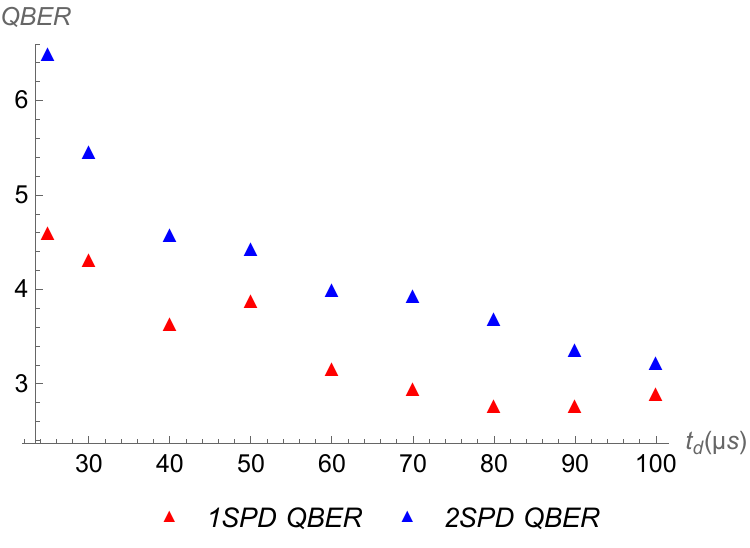}
\caption{QBER when detector efficiency $\eta=0.20$.}
\end{subfigure}
\caption{Comparing the key rates with single detector and dual detectors on the data line for distance $L=120$ km between Alice and Bob.}
\end{figure}

Further, error correction is required based on the QBER to ensure that Alice and Bob generate identical keys. The disclosed rate (\textit{DR}) represents the percentage of the raw key shared over the classical channel used for QBER estimation. We consider \textit{DR} of 10\%, which is statistically sufficient to detect the presence of an eavesdropper by analyzing a randomly selected subset of bits for QBER estimation. We employ the low-density parity-check scheme for error correction, which requires some information exchange over the classical channel. After error correction, privacy amplification is performed to derive the final secure key. This step eliminates any information potentially leaked during error correction via the classical channel. Privacy amplification involves compressing the error-corrected key into a short, completely random bit string. The compression ratio (\textit{CR}) determines the degree of shortening applied to the key during this process. The final secure key rate can be obtained such as \textit{SKR}$\times$(1-\textit{DR}$)\times$(1-\textit{CR}) \cite{Malpani2024}. With a secure \textit{CR}=90\%, we obtain a maximum increase in secure key rates at low $t_d$ as shown in Table 1 using dual SPDs. We observe an increase in the secure key rates of 80\%, 60\%, and 50\% at distances $L=80,100$ and $120$ km. \begin{table}[ht]
\small
\begin{center}
\caption{Increase in secure key rates at varied distance $L$ using dual detectors.}

\begin{tabular}{|c|c|c|c|c|c|c|}

\hline  
 $L$ (km) & \begin{tabular}{c}
     80  \\
       $(\eta=0.15)$ \\
       $(t_d=15\mu s)$ 
 \end{tabular} & \begin{tabular}{c}
     80  \\
       $(\eta=0.20)$ \\
       $(t_d=15\mu s)$ 
 \end{tabular} & \begin{tabular}{c}
     100 \\
       $(\eta=0.15)$ \\
       $(t_d=20\mu s)$ 
 \end{tabular} & \begin{tabular}{c}
     100  \\
       $(\eta=0.20)$ \\
       $(t_d=20\mu s)$ 
 \end{tabular} & \begin{tabular}{c}
     120  \\
       $(\eta=0.15)$ \\
       $(t_d=25\mu s)$ 
 \end{tabular} & \begin{tabular}{c}
     120  \\
       $(\eta=0.20)$ \\
       $(t_d=25\mu s)$ 
 \end{tabular} \\ 
 \hline
  \begin{tabular}{c}
     1 SPD \\
     rates (kbps)
 \end{tabular}& 2.1 & 2.4 & 1.4 & 1.8 & 1 & 1.2 \\
 \hline
  \begin{tabular}{c}
     2 SPD \\
       rates (kbps)
 \end{tabular}& 3.7 & 4.3 & 2.3 & 2.9 & 1.4 & 1.8 \\
  \hline
\end{tabular}

\label{table 16}
\end{center}
\end{table}  
\begin{figure}
    \centering
\includegraphics[width=0.99\linewidth]{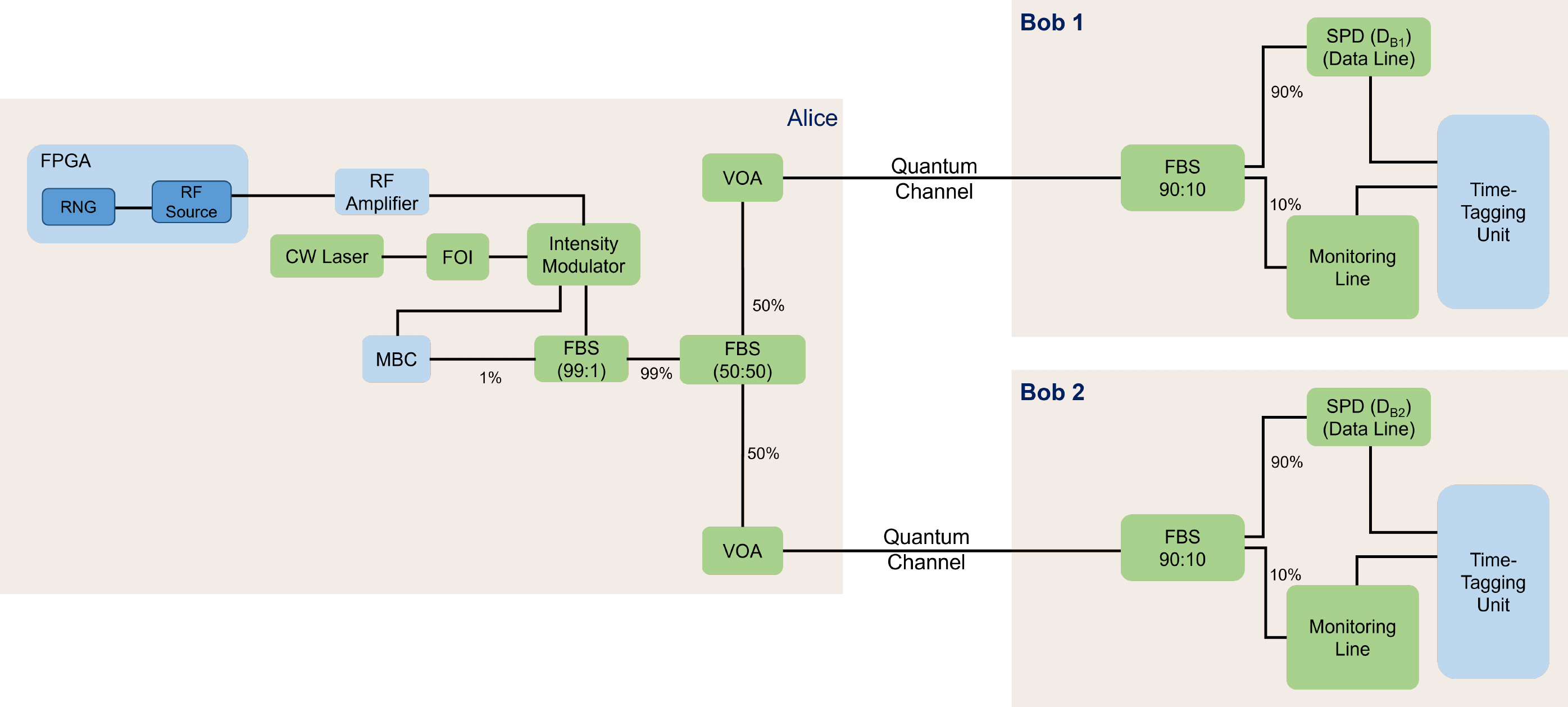}
    \caption{Concurrent COW QKD: Alice module and dual Bob modules}
    \label{fig:enter-label}
\end{figure}

\subsection{Point-to-Multipoint extension}

\begin{figure}[ht!]
\centering
\begin{subfigure}{0.63\linewidth}
\includegraphics[width=\linewidth]{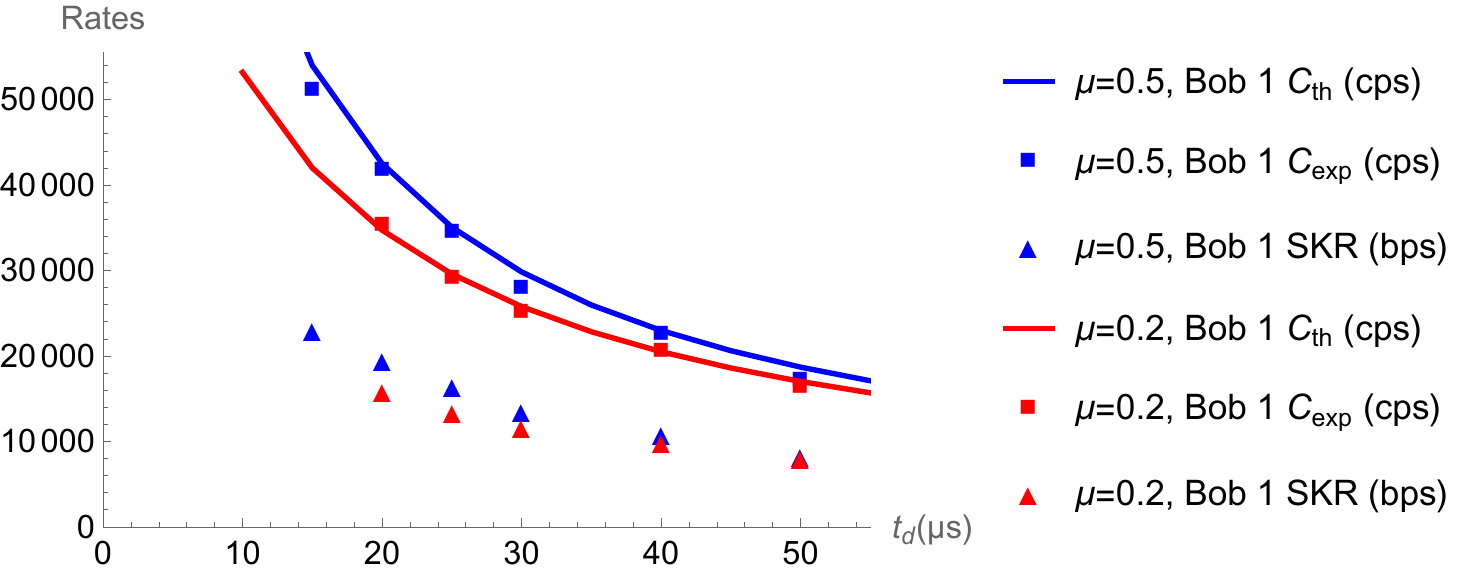}
\caption{Rates of Bob1 module.}
\end{subfigure}
\begin{subfigure}{0.359\linewidth}
\includegraphics[width=\linewidth]{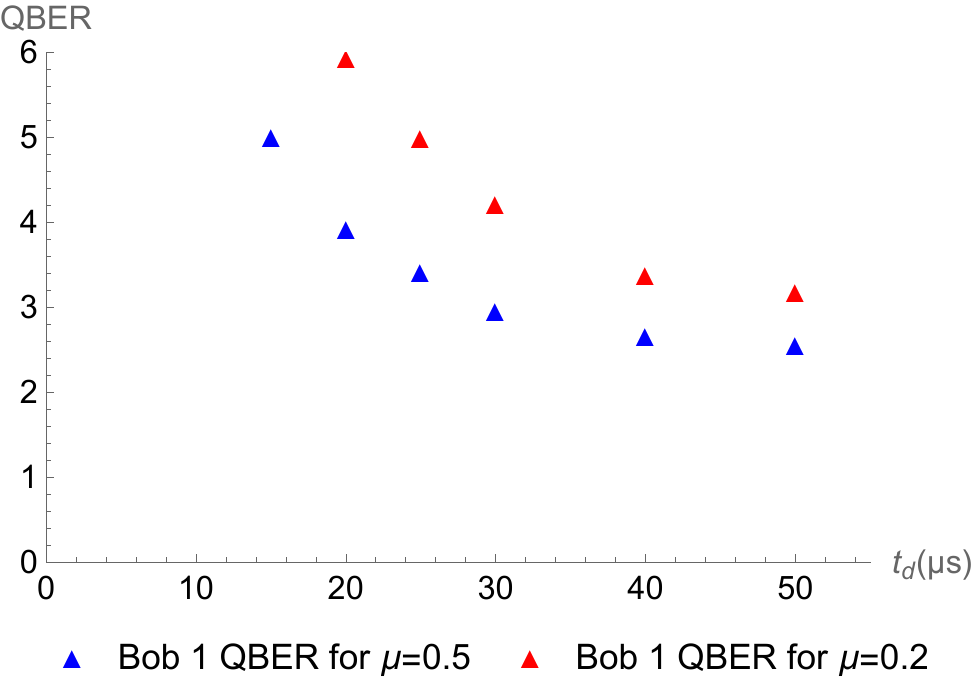}
\caption{QBER at Bob1 module.}
\end{subfigure}
\begin{subfigure}{0.63\linewidth}
\includegraphics[width=\linewidth]{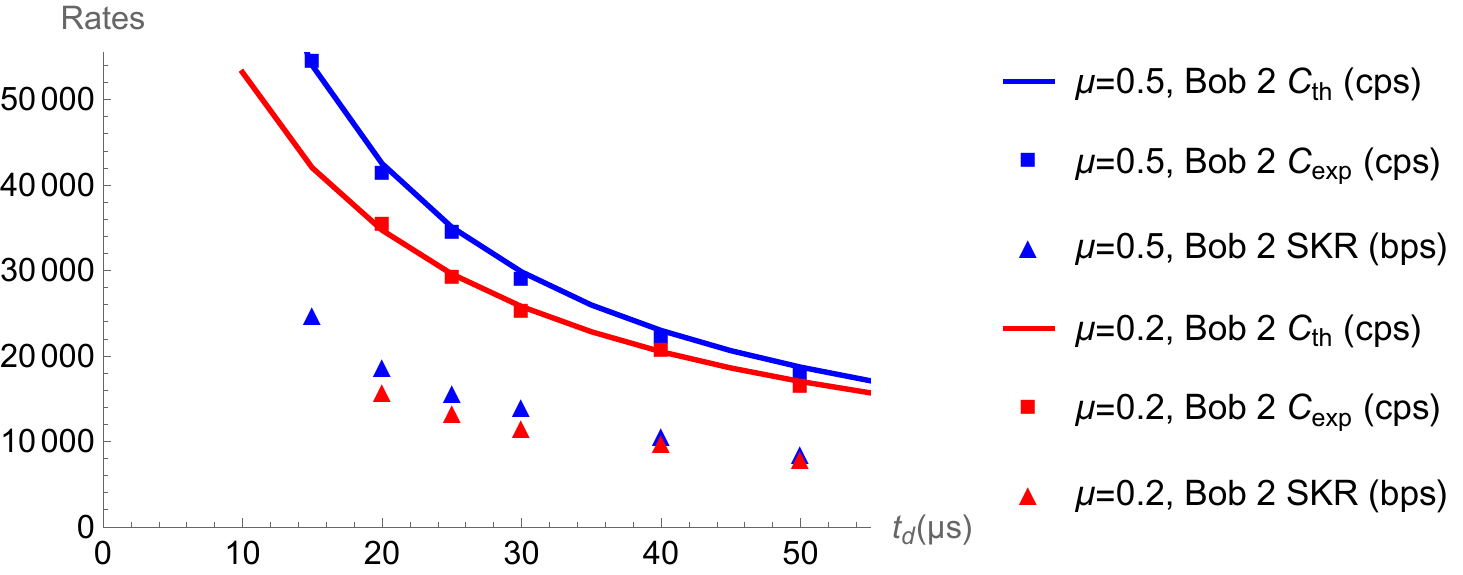}
\caption{Rates at Bob2 module.}
\end{subfigure}
\begin{subfigure}{0.359\linewidth}
\includegraphics[width=\linewidth]{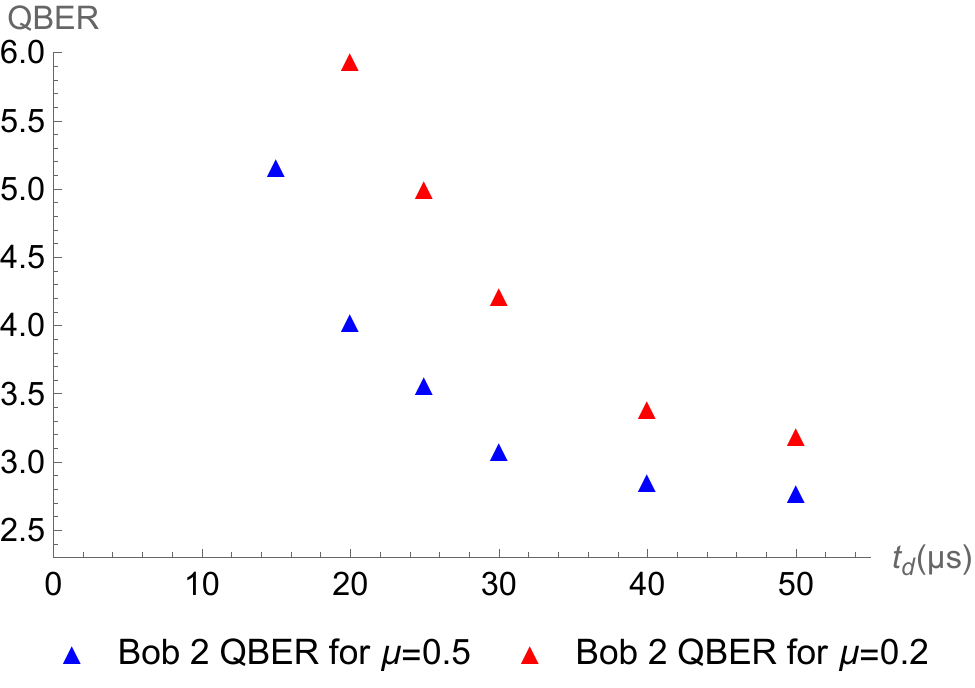}
\caption{QBER at Bob2 module.}
\end{subfigure}
\caption{Comparing the key rates and QBER for Bob 1 module and Bob 2 module for distance $L=100$ km between Alice and Bob 1 (Bob 2) for $\eta=0.2$ with $\mu=0.5$ and $\mu=0.2$. For $\mu=0.2$ we observe lower throughput within the QBER threshold, and we only measure $t_d=20\muup s-50\muup s$ where key rates are in the range of 1 kbps.}
\end{figure}

To increase the secret key transmission between three users (Alice $\to$ (Bob~1, Bob~2)), we introduce an Alice module between the dual Bob modules, as seen in Fig. 7 \cite{Townsend1997,PhysRevLett.119.150501}. After modulating the signal, Alice can create two channels by placing a 50:50 BS. The signal is still classical, and two sets of variable optical attenuators are implemented to create two sets of the quantum signals. These signals are sent through two channels to Bob 1 and Bob 2. \textcolor{black}{Since the optical signal is split prior to attenuation at Alice's end (FBS (50:50) prior to VOAs as depicted in Fig. 7), and each arm is independently attenuated to the desired mean photon number $\mu$, the interference visibility remains unaffected; consequently, the visibility performance in the dual Bob configuration remains equivalent to that of a single Bob setup. The usual interferometer and SPDs setup can be placed in the 10\% arm independently, as shown in Fig. 7, to introduce the standard phase-visibility check \cite{Damien}.} To preserve the fundamental security of the protocol (Two concurrent QKD sessions) and to simplify the entire post-processing procedure, Alice creates a key $k_{A1}$ ($k_{A2}$) separately with Bob 1 (Bob 2). If $|k_{A1}|\ne|k_{A2}|$, then $||k_{A1}|-|k_{A2}||$ bits from the longer key are removed by Alice to ensure $|k_{A1}|=|k_{A2}|$. Alice then communicates $k_{A1} \oplus k_{A2}$ (OTP encrypted) on the classical channel to both Bob 1 and Bob 2. Bob 1 (Bob 2) obtains $k_{A2} = k_{A1} \oplus k_{A1} \oplus k_{A2}$ ($k_{A1} = k_{A2} \oplus k_{A1} \oplus k_{A2}$) using his private copy of $k_{A1}$ $(k_{A2})$.  Bob 1 and Bob 2 obtain the final secure key $k_{A12}$ utilizing \{$k_{A1}$,$k_{A2}$\} after further privacy amplification, discarding one of them to ensure OTP security.

Using the experimental setup shown in Fig. 7 for a distance $L=100$ km between Alice $\to$ (Bob~1, Bob~2) (same distances), the $C_{th}$ (cps), $C_{exp}$ (cps), and \textit{SKR} for $k_{A1}$ (Fig. 8(a)) and $k_{A2}$ (Fig. 8(c)) (bps) are shown in Fig. 8 for varying mean photon numbers $\mu=0.5$ and $\mu=0.2$ with $\eta=0.2$. The experimental rates $C_{exp}$ and \textit{SKR} are compared for the Bob 1 module and the Bob 2 module. Below the QBER threshold of $5\%$, both Bob 1 and Bob 2 give a lower throughput of 1.2 kbps for $\mu=0.2$, while $\mu=0.5$ gives a throughput of 1.8 kbps. Bob 1 module and the Bob 2 module rates seem comparable with minimal QBER, and for $\mu=0.5$, the rates are similar to the single SPD rates in a typical single receiver COW implementation in Fig. 5 (c), (d) (1SPD). We note that while we experimented with $L=100$ km between Alice and Bob 1 (Bob 2) for comparison with the results from the existing section for $\mu=0.5$, the effective combined distance bound between Bob 1 and Bob 2 modules will be less than 100 km for secure QKD communication, as can be seen in a further section. The potential information gain for an eavesdropper increases if she can simultaneously access both channels carrying quantum signals, since exploiting correlations between them can enhance her ability to infer the secret key \cite{bian2023highratepointtomultipointquantumkey}. We explicitly show this by considering the collective beam-splitting attack further for the COW protocol \cite{Branciard_2008}. \textcolor{black}{Crucially, an eavesdropper attacking both channels could gain information on both keys $k_{A1}$ and $k_{A2}$. In our dual Bob model with two channels, total information leakage to Eve exceeds that of a single channel. During further privacy amplification, we should therefore compress the combined key $k_{A12}$ by at least this amount. Therefore, a sufficiently large compression ratio should be chosen to eliminate this worst-case leakage, ensuring that the final shared key $k_{A12}$ is secure.}

\subsubsection{Collective Beam-splitting attack for COW}
In the typical Beam-splitting attack (BSA) \cite{Branciard_2008}, the lossy channel is replaced by an ideal lossless channel, and a beam splitter is inserted between Alice and Bob that diverts the fraction $t_E=1-t_{B}$ of the optical power to her quantum memory while forwarding the remaining fraction $t_{B}$ to the legitimate receiver Bob. Because the output mode forwarded to Bob exactly reproduces the expected lossy mode, this attack introduces no errors in the data line (hence QBER $=0$) and preserves full coherence (ideal visibility); thus, this attack is undetectable by the usual parameter estimation based on QBER and visibility.

For the typical COW protocol \cite{Branciard_2008}, the eavesdropper's retained local amplitude is characterized by
\begin{equation}
\gamma_E \;=\; e^{-\mu t_{E}},
\label{eq:gammaE}
\end{equation}
with $\mu$ the mean photon number of non-empty pulses and $t_B=10^{\left(\frac{-\alpha_d L}{10}\right)}$ is the channel transmissivity. Under the BSA, when Alice encodes a bit using the COW protocol, the two relevant states available to Eve can be taken as the two-mode coherent states $|\psi_0\rangle_E = |\sqrt{\mu_E}\rangle\otimes|0\rangle, 
|\psi_1\rangle_E = |0\rangle\otimes|\sqrt{\mu_E}\rangle,$ for bits $0$, $1$, respectively corresponding to the cases where the non-empty pulse is in the earlier or the later time slot for Eve's states with $\mu_E=\mu t_E$. Their inner product factorizes to give $\langle\psi_0|\psi_1\rangle_E = \gamma_E$.
Further, utilizing this, the Holevo information that the eavesdropper can obtain about Alice's bits $\chi_{AE}$ (and equivalently Holevo information related to Bob's bits for the BSA case $\chi_{BE}$) reduces to
\begin{equation}
\chi_{AE} \;=\;\chi_{BE}\;= \; h\ \!\bigg(\frac{1-\gamma_E}{2}\bigg),
\label{eq:chiCOW}
\end{equation}
where $h(x)=-x\log_2 x - (1-x)\log_2(1-x)$ is the binary entropy. The Holevo quantity \(\chi_{AE}(\chi_{BE})\) upper bounds the classical mutual information that Eve can obtain about Alice's (Bob's) classical bit string if she is allowed to perform arbitrary collective quantum measurements on her quantum memory and optimal (coherent) classical post-processing. More precisely, for many independent uses of the channel and collective (but identical) attack strategies, the accessible information per signal is asymptotically limited by the Holevo information; this is why \(\chi\) is the appropriate measure for collective attacks in the asymptotic Devetak–Winter secure key rate analysis \cite{rspa.2004.1372}. In the trusted-device scenario (detector efficiency $\eta$ fixed), the corresponding Devetak--Winter secret key rate per pulse \cite{rspa.2004.1372} for a single Bob under this attack is
\begin{equation}
r_{\mathrm{B}}(\mu,t_B) \;=\; \frac{1}{2}\Big(1-e^{-\mu t_B \eta}\Big)\Big[1-\chi^{\mathrm{COW}}_{E}(\mu,t_B)\Big],
\label{eq:rCOW}
\end{equation} with Holevo information $\chi^{\mathrm{COW}}_{E}=\;\chi_{BE}$. Eqs. \eqref{eq:gammaE}--\eqref{eq:rCOW} are used to compute the single-Bob bounds.

We now adapt the BSA model to the point-to-multipoint (Alice $\to$ (Bob~1, Bob~2)) topology used in our experiment. Operationally, Alice splits the modulated optical signal and delivers (attenuated) states to Bob~1 and Bob~2. A conservative/worst-case security assumption is that the eavesdropper can coherently access and store the modes lost from both channels of the broadcast (for example, by simultaneously attacking both physical fibers). Under this assumption, the eavesdropper may correlate her probes across the two channels and thus may achieve greater information than in the independent, uncorrelated case. To obtain a tight and simple bound, we therefore impose
\begin{equation}
\chi^{\mathrm{COW}}_{E}(\mu,t_B)\;=\;\chi_{\,\mathrm{Bob1}~E} + \chi_{\,\mathrm{Bob2}~E} = 2 \chi_{BE},
\label{eq:equal_chi}
\end{equation}
i.e., eavesdropper's Holevo information is combined and doubled. This models the worst case where (i) the two channels are identical (same loss, same detector efficiency) and (ii) the eavesdropper exploits all the correlations available between the two channels. Taking Eq. \eqref{eq:equal_chi} is conservative because any real, imperfect asymmetry between the channels or any inability of the eavesdropper to coherently correlate her stored modes would typically reduce her joint information; thus, it gives a tighter (i.e., more pessimistic) bound on the achievable SKR for the network. Under this symmetric dual-Bob assumption, each receiver individually has the per-pulse rate given by  Eq. \eqref{eq:rCOW} with  Eq. \eqref{eq:equal_chi}. \textcolor{black}{During a final privacy amplification, the key is to be shortened by at least Eve’s combined leakage from both channels. From Eq.~\eqref{eq:equal_chi}, the compression must therefore account for the total Holevo information $\chi^{\mathrm{COW}}_{E}(\mu,t_B)$, so that the final key $k_{A12}$ retains only the fraction $\big[1-\chi^{\mathrm{COW}}_{E}(\mu,t_B)\big]$ that is information-theoretically secure.}

\begin{figure}[htb]
  \centering
  \includegraphics[width=0.99\textwidth]{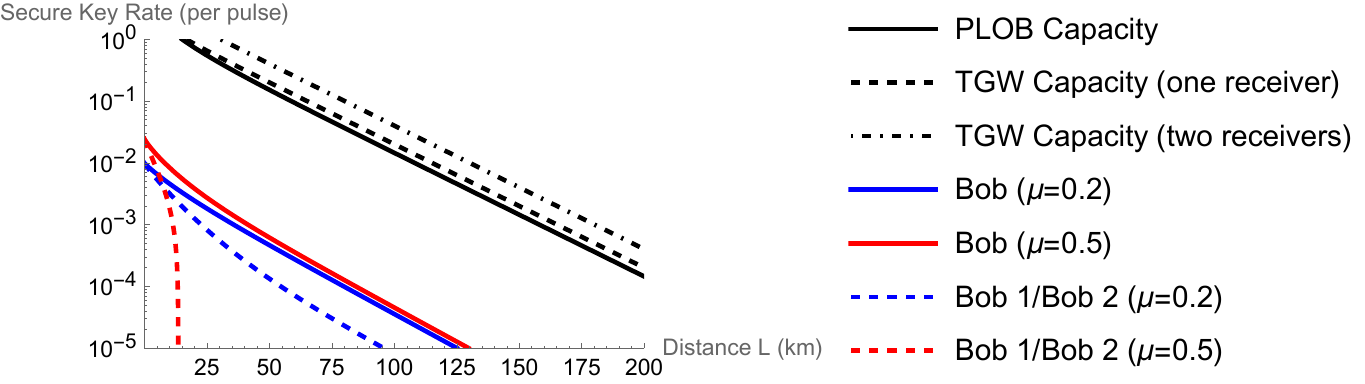}
  \caption{Secure key rates per pulse as a function of distance $L$. Curves labelled 'Bob ($\mu=0.2,0.5$)' are single-receiver secret key rates computed from Eq.~\eqref{eq:rCOW}; 'Bob 1/Bob 2' denotes individual receiver rates for the two-receiver scenario. 'PLOB Capacity' gives the exact secret-key capacity of a pure-loss optical channel \cite{Pirandola2017}. 'TGW Capacity (one receiver)' and 'TGW Capacity (two receivers combined)' are the ideal unconstrained capacity bounds from the pure-loss bosonic broadcast channel for single-receiver and two-receivers, respectively \cite{PhysRevLett.119.150501}.}
  \label{fig:plotv2}
\end{figure}

Fig.~\ref{fig:plotv2} gives the numerical results for key rates $r_{\mathrm{B}}(\mu,t_B)$ from this analysis. The rates for single-Bob and dual-Bob cases were generated by evaluating Eq.~\eqref{eq:rCOW} as functions of distance $L$, using the experimental parameters employed in Sec. 2.1 for $\eta=0.2$. The two mean photon numbers $\mu=0.5$ and $\mu=0.2$ were used for the plots. The single-Bob rates for $\mu=0.5$ and $\mu=0.2$ reduce for increasing $L$ nearly the same way. For the dual-Bob case $\mu=0.2$ curve outperforms in $L$ compared to the $\mu=0.5$ in Fig.~\ref{fig:plotv2}. For the larger intensity $\mu=0.5$, the eavesdropper receives a stronger reflected mode (larger mean photon number in her retained mode), increasing $\chi^{\mathrm{COW}}_{E}$ via Eq.~\eqref{eq:equal_chi}. As a consequence, the factor $[1-\chi^{\mathrm{COW}}_{E}]$ in Eq.~\eqref{eq:rCOW} decreases, and the secure key rate falls off rapidly with distance $L$; thus $\mu=0.5$ is suboptimal for long links with dual-Bob. For the smaller intensity $\mu=0.2$, the eavesdropper's retained mode contains fewer photons on average; consequently, $\chi^{\mathrm{COW}}_{E}$ is smaller and the secure fraction $[1-\chi^{\mathrm{COW}}_{E}]$ remains significantly larger at long distances, yielding superior secure key rate at long-distance. To place these dual-Bob bounds in a general context and to show that combined key rates generally increase, we also plot the ideal information-theoretic capacity limits for the pure-loss bosonic broadcast channel \cite{PhysRevLett.119.150501} and the PLOB bound \cite{Pirandola2017}. For a 1-to-2 pure-loss broadcast channel, the TGW capacity is $\le\; -\log_2\big(1 - 2t_B)$ with receivers' transmittance $t_B$ (receivers with the same transmittance), in bits per pulse, which upper bounds the sum of secure key rates between Alice and each receiver \cite{PhysRevLett.119.150501}. Similarly, the TGW upper bounds for individual secure key rates between Alice and each receiver are $\le\; -\log_2\big([1-t_B]/[1-2t_B])$, which is lower.  For a 1-to-1 channel and 1-to-2 channel, the PLOB result gives the tighter secret-key capacity as $\le\; -\log_2\big(1-t_B)$ \cite{PhysRevA.96.032318}.  The `TGW Capacity (two receivers combined)', `TGW Capacity (one receiver)', and `PLOB Capacity' curves in Fig.~\ref{fig:plotv2} are plotted based on this.

\subsubsection{Extension to a network}
Assume we are given a set of $N$ parties to generate a key by utilizing the description in the above section. One should initially design an optimised quantum network for the $N$ parties, where each segment (Bob 1 - Alice - Bob 2) should share a vertex with at least one other segment (For example, see Fig. 10 for $N=5$), based on various practical constraints. Initially, the prescription described in the above section will be executed in each segment \{1,2,3\} and \{3,4,5\}. The \textit{inter-segmental} keys generated within a segment can be utilized to generate a final key. Suppose $k_{123}$ ($k_{A12}$ in Sec. 2.2) and $k_{345}$ are the keys derived from two consecutive segments concurrently in Fig. 10. $k_{123} \oplus k_{345}$ can be shared over the classical channel to all the parties and procedure described in above section can be used to reconcile the secure key. The secret key rate finally produced from {$k_{123},k_{345}$} depends only on the longest nearest-neighbor distance within all segments ($d_{1,2}$ in Fig. 10), rather than linearly depending on the network size ($d_{1,5}$ in Fig. 10) \cite{PhysRevApplied.14.024010}. The network topology typically determines how many segments are needed.

\begin{figure}
    \centering
\includegraphics[width=0.5\linewidth]{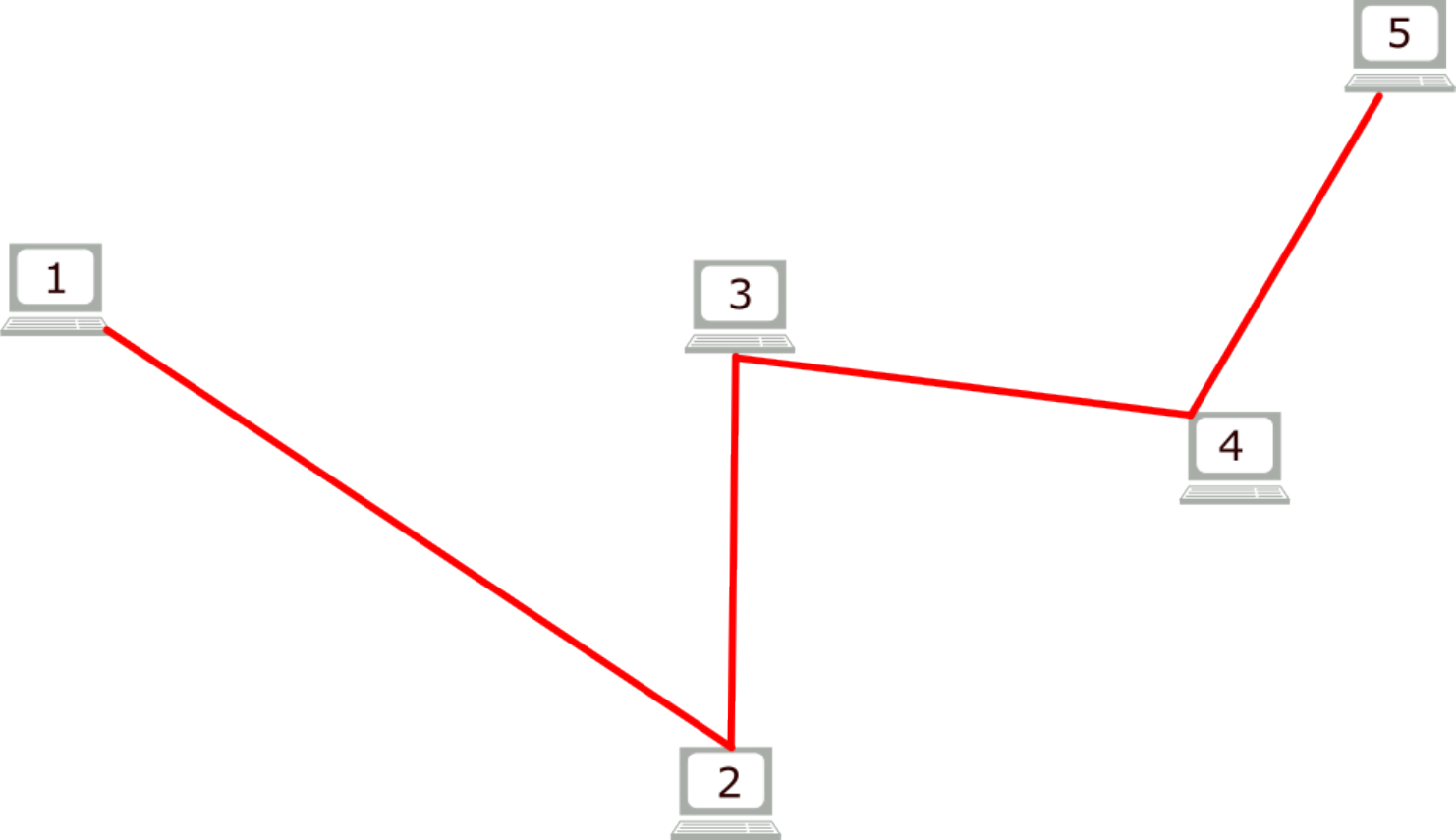}
    \caption{A simple network having $N=5$ parties and $n=2$ concurrent COW implementation between $\{1,2,3\}$ and $\{3,4,5\}$. Vertices 2 and 4 act as Alice, with 1, 3 being Bob 1 and 3, 5 being Bob 2. The key rate depends only on the longest nearest-neighbor distance, in this example $d_{1,2}$, rather than linearly with the network size (in this case $d_{1,5}$).}
    \label{fig:enter-label}
\end{figure}

\section{Conclusion}
In conclusion, we have demonstrated methods to enhance the secret key rates and transmission for multiple users using the COW-QKD protocol without altering its fundamental framework. We experimentally showed the improvement of secure key rates by leveraging dual SPDs on the receiver's data line and the transmission to three users by introducing an additional receiver module. The results confirm that integrating dual SPDs improves the secure key rates while maintaining QBER within acceptable thresholds for distances. These results highlight that our approach can be generalized to other time-bin encoding-based QKD protocols \cite{10.1063/5.0232085,montaut2025progress,singh2025photonicquantuminformationtimebins,xavier2025energy}, especially in regimes where detector dead time limits the sifted key rate \cite{Maximilian2025}. When concerned with COW protocol, especially, this effect becomes even more pronounced when a $50:50$ (data line : monitoring line) beam splitter is employed for Bob’s passive basis choice, instead of the conventional $90:10$ design \cite{PhysRevApplied.18.064053}. Further, the high-dimensional time-bin COW-QKD protocol is a promising new direction \cite{Sulimany2025High}. A 32-dimensional time-bin COW-QKD protocol using a standard two-detector setup was experimentally demonstrated. By permuting the time bins (with no hardware changes), they achieved about a twofold increase in the asymptotic secure key rate compared to the standard COW protocol.

Extending the typical COW-QKD approach with dual receivers, we propose a scalable architecture that generalizes key sharing across multiple parties and enables straightforward scaling of secure communications. Under a conservative collective beam-splitting attack model \cite{Branciard_2008}, our numerical evaluation shows that operating at a lower mean photon number ($\mu=0.2$) yields substantially better long-distance secure key rates than $\mu=0.5$ for the experimental parameter regime considered. Although the dual-receiver extension can increase aggregate secret throughput, achievable rates remain ultimately constrained by the eavesdropper’s capabilities under collective and coherent attacks; therefore, practical point-to-multipoint deployments must carefully optimise source intensity and explicitly account for stronger attack models to obtain meaningful security margins. While COW-QKD is inherently robust against individual attacks \cite{Damien}, theoretical bounds must be re-evaluated in the presence of zero-error and other coherent attacks \cite{PhysRevLett.125.260510}, particularly when keys are generated as described in Sec. 2.2. Related studies include recent simulations of multi-Bob networks for BB84 under individual attacks \cite{10.1117/12.3027618}. Also, proposals combining COW with the novel twin-field QKD to produce high-rate conference keys for multiple users are interesting \cite{cao2021coherent}. Twin-field approaches typically require a more complex, measurement-device-independent receiver architecture \cite{Lucamarini2018}, and we have recently evaluated point-to-multipoint secret key rates for twin-field multi-party agreements \cite{Abhignan:25}. Finally, our prior optical simulations of hacking attempts on two-party COW \cite{Abhignan_2024}, such as backflash attacks \cite{singh2025backflash}, show that experimental hacking must be re-examined carefully for multi-party and dual-receiver deployments, and such practical attack analyses should accompany any real-world network deployments.

\section*{Acknowledgements}
The authors sincerely thank Prof. Stefano Pirandola for his kind personal communication, which pointed out the correct interpretation of the TGW and PLOB bounds, and for sharing the relevant reference on the extension to the 1-to-2 scenario \cite{PhysRevA.96.032318}. The authors also thank Dr. Ashutosh Singh and Dr. R. Srikanth for detailed discussions on the experimental and theoretical aspects of the work as well as for helpful suggestions that strengthened the manuscript.
\\
\\
Data sets generated during the current study are available from the corresponding author on reasonable request.

\bibliographystyle{ieeetr}
\bibliography{article.bib}
\end{document}